\documentclass[a4paper,11pt,dvips]{article}
cm \voffset = -2true cm

\usepackage{graphicx}
\usepackage{amsmath}
\usepackage{amssymb}
\usepackage{latexsym}
\usepackage[colorlinks]{hyperref}
\usepackage{color}
\usepackage{color}
\usepackage{float}
\usepackage{cite}
\usepackage{makeidx}
\usepackage{colortbl}
\usepackage{dsfont}
\newcommand{\bra}{\begin{array}}
\newcommand{\era}{\end{array}}
\newcommand{\beq}{\begin{equation}}
\newcommand{\eeq}{\end{equation}}
\newcommand{\beqar}{\begin{eqnarray}}
\newcommand{\eeqar}{\end{eqnarray}}

\def\BC{\bb C}
\def\_\BC{\bbi C}

\def\( {\left(}
   \def\) {\right)}
\def\[ {\left[}
\def\] {\right]}
\def\no2 {{\textstyle{n\over 2}}}

\def\dag {{\dagger}}

\newcommand{\lb}{\label}

\begin{document}

\begin{titlepage}
\setcounter{page}{1}
\renewcommand{\thefootnote}{\fnsymbol{footnote}}

\begin{flushright}
\end{flushright}

\vspace{5mm}
\begin{center}

{\Large \bf {{Electron Scattering in Gapped
Graphene Quantum Dots}}}

\vspace{5mm}

{\bf Abdelhadi Belouad}$^{a}$, {\bf Youness Zahidi}$^{b}$,  {\bf
Ahmed Jellal\footnote{\sf 
a.jellal@ucd.ac.ma}}$^{a,c}$, {\bf Hocine Bahlouli}$^{c,d}$

\vspace{5mm}

{$^{a}$\em Theoretical Physics Group,  
Faculty of Sciences, Choua\"ib Doukkali University},\\
{\em PO Box 20, 24000 El Jadida, Morocco}

{$^{b}$\em MATIC laboratory, FPK, Hassan 1 University, Khouribga,
Morocco}

{$^c$\em Saudi Center for Theoretical Physics, Dhahran, Saudi
Arabia}

{$^d$\em Physics Department,  King Fahd University
of Petroleum $\&$ Minerals,\\
Dhahran 31261, Saudi Arabia}


\vspace{3cm}

\begin{abstract}

{Due to Klein tunneling in graphene only quasi-bound states are
realized in graphene quantum dots by electrostatic gating.
Particles in the quasi-bound states are trapped inside the dot for
a finite time and they keep bouncing back and forth till they find
their way out. Here we study the effect of an induced gap on the
scattering problem of Dirac electrons on a circular
electrostatically confined quantum dot. Introducing an energy gap
inside the quantum dot enables us to distinguish three scattering
regimes instead of two in the case of gapless graphene quantum
dot. We will focus on these regimes and analyze the scattering
efficiency as a function of the electron energy, the dot radius
and the energy gap. Moreover, we will discuss how the system
parameters can affect the scattering resonances inside the dot.}


\end{abstract}
\end{center}

\vspace{3cm}
\end{titlepage}

\section{Introduction}

Graphene \cite{1} is a material consisting of a single atomic
layer of carbon in $sp^2$ hybridization. It can be viewed either
as a single layer of graphite or an unrolled nanotube.
Specifically the electronic properties of graphene are
extraordinary. This is why graphene has attracted a lot of
interest in fundamental physics for its possible technological
applications~\cite{1,2,3,4,5}. Graphene can provide a good
platform for the study of the electronic properties of a pure
two-dimensional system. In graphene the quasi-particles
(low-energy excitations) close to the Dirac nodal points behave
like mass-less relativistic Dirac fermions with a linear energy
dispersion. In addition, graphene presents a variety of exotic
electronic properties like electronhole symmetry \cite{2}, Klein
tunneling \cite{51} and anomalous quantum Hall effect \cite{52}.

The equation describing the electronic excitations in graphene is
formally similar to the Dirac equation for massless fermions,
which travel at a speed of the order of $v_F \approx 10^6 m
s^{-1}$ \cite{53,54}.
As a consequence of the pseudo-relativistic dynamics, the massless
Dirac fermions have an additional pseudospin degree of freedom.
That is the perfect transmission through arbitrarily high and wide
rectangular potential barriers or $n-p$ junctions at normal
incidence. Unfortunately, the Dirac fermions cannot be confined by
electro-static potentials. This is due the Klein tunneling effect
\cite{51} and the absence of the gap in the energy spectrum. Thus
the realization of the quantum dots is needed to overcome such
situation. Recently, alternative strategies have been proposed to
confine charged particles by using thin single-layer graphene
strips \cite{55,56} or nonuniform magnetic fields \cite{57}.
Graphene quantum dots \cite{56,58,60} have been recently
extensively discussed theoretically as well as from the
experimental side \cite{601,602,603,604,605,606}.
It have been studied as potential
hosts for spin qubits~\cite{16,19}, single gate-defined dots
\cite{62}. In addition, multiple dots arranged in
corrals~\cite{19} have been used to model the scattering of Dirac
electron waves by impurities or metallic islands placed on a
graphene sheet.

Different experimental methods are available to open a gap in
graphene band structure, called the Dirac gap~\cite{3}. As
demonstrated in the experiment, the maximum energy gap could be
260meV due to the sublattice symmetry breaking \cite{23}. It is
important to note that the value of the energy gap changes by
changing the experimental technique. One of the experimental
methods used to open a gap has been demonstrated by controlling
the structure of the interface between graphene and ruthenium
\cite{64}. Moreover, in graphene grown epitaxially on a SiC
substrate an energy gap has been measured \cite{23}. In addition,
it has been demonstrated theoretically that an energy gap can be
opened by the application of a local strain and/or a chemical
methods \cite{3,67,68,69}.

We study the electron propagation in a circular electrostatically
defined quantum dot in monolayer graphene in the presence of an
energy gap inside the dot. We identify different scattering
regimes depending on the radius, potential and Dirac gap of the
dot as well as the electron energy. Then, we 
determine
the scattering coefficients as well as the radial component of the
corresponding reflected current. Subsequently, we study
the
scattering efficiency Q, which is defined as the scattering cross
section divided by the geometric cross section of a plane Dirac
electron wave hitting on a quantum dot in graphene.  The main characteristics
of these quantities will be studied in terms of
the physical parameter of our system.

The present paper is organized as follows. In section $2$, we
present a theoretical study of propagation wave plane of electron
in a circular quantum dot of monolayer graphene. We give the
solutions of the spinors of the Dirac equation corresponding to
each region of different scattering parameters. We use the
continuity of the wave functions at the boundary of the dot in
order to calculate the scattering coefficients. In section $3$, we
analyze the scattering efficiency, square modulus of the
scattering coefficients and radial component of the far-field. We numerically
discuss our results by giving  different illustrations. Finally, we close
our work by summarizing the main obtained results.

\section{Theoretical model}

For a Dirac electron in a circular electrostatically defined
quantum dot in monolayer graphene 
with gap $\Delta(r)$, the single-valley Hamiltonian, in the unit system
($\hbar=v_F=1$), can be written as
\begin{equation}\lb{eq1}
 H=-i\vec{\nabla}\cdot \vec \sigma+V(r)\mathds{1}+\Delta(r)\sigma_z
\end{equation}
where $\vec \sigma=(\sigma_x, \sigma_y,\sigma_z)$ are the Pauli
matrices {and $\mathds{1}$ is the $2\times 2$ unit
matrix}. The applied bias $V(r)$ and
$\Delta(r)$ are given by
\begin{equation}
 V(r) =  \left \{ \begin{array}{cc} 0, &  r>R
\\ V, &   r\leq R
\end{array} \right., \qquad
 \label{eq:Uform} \Delta(r) =  \left
\{ \begin{array}{cc} 0, &  r>R
\\ \Delta, &  r\leq R
\end{array} \right.
\end{equation}
and
$R$ is the quantum dot radius  as depicted schematically in Figure \ref{Fig.1}:

\begin{figure}[!ht]
\centering
\includegraphics[width=10cm]{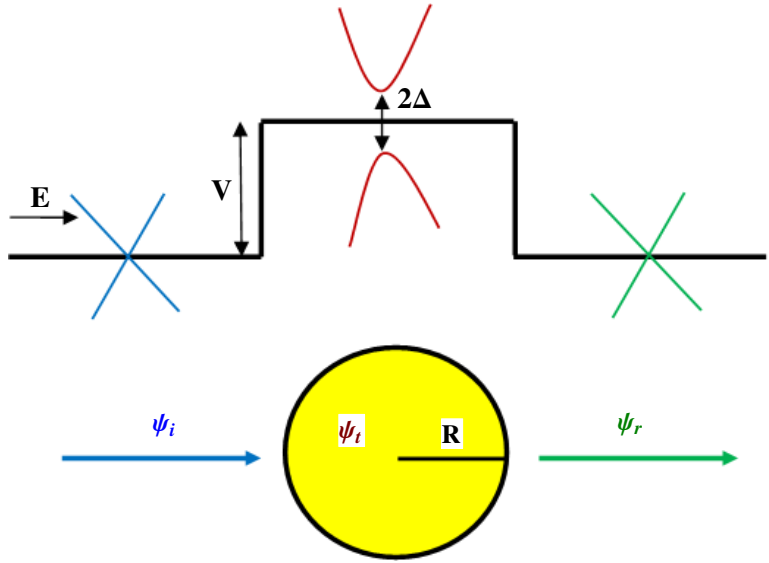}
\caption{\sf{ Sketch of Dirac electron scattering for a low energy
at a graphene quantum dot in the presence of a gap $\Delta$. The
quantum dots are defined electrostatically by applying a constant
bias $V$. For $E < V-\Delta$, the incident $\psi_i$ and reflected
$\psi_r$ electron waves reside in the conduction band, while the
transmitted $\psi_t$ wave inside the dot corresponds to a state in
the valence band.}} \label{Fig.1}
\end{figure}

The geometry presented in Figure \ref{Fig.1} suggests to map the system Hamiltonian
\eqref{eq1} in the polar coordinates $(r, \phi)$
as
\begin{equation}
\label{eq2} H =
\begin{pmatrix}
V_+ & e^{-i\phi}\left(-i\frac{\partial}{\partial r}-\frac{1}{r}\frac{\partial}{\partial \phi}\right) \\
e^{i\phi}\left(-i\frac{\partial}{\partial r}+\frac{1}{r}\frac{\partial}{\partial \phi}\right) & V_- \\
\end{pmatrix}
\end{equation}
where we have defined $V_{\pm}=V\pm\Delta$. One can easily check that  $H$ commutes
with the total 
momentum operator
$J_z=L_z+\frac{1}{2}\sigma_z$, as consequence the  
 eigenspinors can be
chosen to be eigenstates of $J_z$ and therefore  they are separated into
radial $R^{\pm}(r)$ and angular $\chi(\phi^{\pm})$ parts
\begin{equation}
\label{eq3} \psi_{m}(r,\phi)=\begin{pmatrix}
R_{m}^{+}(r)\chi_{m}^{+}(\phi)\\
R_{m+1}^{-}(r)\chi_{m+1}^{-}(\phi)\\
\end{pmatrix}
\end{equation}
with
the eigensates 
\begin{equation}
\label{eq3}
\chi^+(\phi)=\frac{e^{im\phi}}{\sqrt{2\pi}}\begin{pmatrix}
1\\
0\\
\end{pmatrix},
\qquad
\chi^-(\phi)=\frac{e^{i(m+1)\phi}}{\sqrt{2\pi}}\begin{pmatrix}
0\\
1\\
\end{pmatrix}
\end{equation}
and $m=0, \pm 1, \pm 2, \cdots$, being the orbital angular momentum
quantum number.

In order to get the solutions of the energy spectrum, we have to
solve the eigenvalue problem $H\psi_m(r,\phi)=E\psi_m(r,\phi)$ by
considering two regions according to Figure \ref{Fig.1}: outside
($r>R$) and inside ($r\leq R$) the quantum dot. Thus we have an
incident wave propagation in the $x$ direction, the reflected wave
is an outgoing wave and a transmitted wave inside the quantum dot. Indeed,
for $r>R$, we show that the radial parts $R_{m}^{+}(r)$ and $R_{m+1}^{-}(r)$
satisfy two coupled differential equations
\begin{eqnarray}
&&-i\frac{\partial}{\partial
r}R_{m}^{+}(r)+i\frac{m}{r}R_{m}^{+}(r)=ER_{m+1}^{-}(r)\\&&
-i\frac{\partial}{\partial
r}R_{m+1}^{-}(r)-i\frac{m+1}{r}R_{m+1}^{-}(r)=ER_{m}^{+}(r)
\end{eqnarray}
giving rise the second differential equation for $R_{m}^{+}(r)$
\begin{equation}
\left(r^2\frac{\partial^2}{\partial^2 r}+r\frac{\partial}{\partial
r}+r^2E^2-m^2\right)R_{m}^{+}(r)=0
\end{equation}
which having the Bessel functions $J_{m}(Er)$ as solution.
Recalling that, we can expand the incident plane wave as
\begin{equation}\lb{eq.9}
\psi_i(r,\phi)=\frac{e^{ikx}}{\sqrt{2}}\begin{pmatrix}
1\\
1\\
\end{pmatrix}=\frac{1}{\sqrt{2}}e^{ikr \cos\phi}
\begin{pmatrix}
1\\
1\\
\end{pmatrix}
=\frac{1}{\sqrt{2}}\sum_m i^{m}J_{m}(kr) e^{im\phi}
\begin{pmatrix}
1\\
1\\
\end{pmatrix}.
\end{equation}
Using \eqref{eq3},
to write the incident spinor as 
\begin{equation}
\psi_i(r,\phi)= \sqrt{\pi} \sum_m
i^{m+1}\left[-iJ_{m}(kr)\frac{1}{\sqrt{2\pi}} e^{im\phi}
\begin{pmatrix}
1\\
0\\
\end{pmatrix}
+J_{m+1}(kr)\frac{1}{\sqrt{2\pi}} e^{i(m+1)\phi}
\begin{pmatrix}
0\\
1\\
\end{pmatrix}
\right]
\end{equation}
as well as the reflected one
\begin{equation}
\psi_r(r,\phi)= \sqrt{\pi}\sum_m
i^{m+1}a_{m}\left[-iH^{(1)}_{m}(kr)\frac{1}{\sqrt{2\pi}}
e^{im\phi}
\begin{pmatrix}
1\\
0\\
\end{pmatrix}
+H^{(1)}_{m+1}(kr)\frac{1}{\sqrt{2\pi}} e^{i(m+1)\phi}
\begin{pmatrix}
0\\
1\\
\end{pmatrix}
\right]
\end{equation}
where $H^{(1)}_{m}(kr)$ are the Hankel function of the first kind,
$a_m$ are the scattering coefficients and the wave number $k = E$.
Now
 for the second case
$r\leq R$,
we have
\begin{eqnarray}
&&-i\left(\frac{\partial}{\partial
r}-\frac{m}{r}\right)R_{m}^{+}(r)=(E-V_{-})R_{m+1}^{-}(r)\\&&
-i\left(\frac{\partial}{\partial
r}+\frac{m+1}{r}\right)R_{m+1}^{-}(r)=(E-V_{+})R_{m}^{+}(r)
\end{eqnarray}
which allow to obtain
\begin{equation}
\left(r^2\frac{\partial^2}{\partial^2 r}+r\frac{\partial}{\partial
r}+r^2\eta^2-m^2 \right)R_{m}^{+}(r)=0
\end{equation}
where we have set $\eta^2=
(E-V)^2-\Delta^2$. This gives the transmitted spinor as
\begin{equation}
\psi_{t}(r,\phi)=\sqrt{\pi} \sum_m i^{m+1}b_{m}\left[-iJ_{m}(\eta
r)\frac{1}{\sqrt{2\pi}} e^{im\phi}
\begin{pmatrix}
1\\
0\\
\end{pmatrix}
+\mu J_{m+1}(\eta r)\frac{1}{\sqrt{2\pi}} e^{i(m+1)\phi}
\begin{pmatrix}
0\\
1\\
\end{pmatrix}
\right]
\end{equation}
with $\mu= \sqrt{\frac{E-V_+}{E-V_-}}$ and $b_m$ are the
scattering coefficients. Later on, we will see the above results can be used to
to study the scattering of Dirac electrons in our system.

\section{Scattering problem}

To study the scattering problem of our system, we need
first to
 determine the scattering coefficients $a_m$ and $b_m$.
This can be done by requiring the eigenspinors continuity at the
boundary $r=R$,
$\psi_i(R)+\psi_r(R)=\psi_t(R)$, to end up with
two conditions
\begin{eqnarray}
&&J_{m}(kR)+a_{m}H^{(1)}(kR)=b_{m}J_{m}(\eta R)
\\&&
J_{m+1}(kR)+a_{m}H_{m+1}^{(1)}(kR)=\mu b_{m}J_{m+1}(\eta R)
\end{eqnarray}
which can be solved to obtain $a_m$ and $b_m$
\begin {equation}\lb{am}
a_{m}=\frac{-J_{m}(\eta R)J_{m+1}(kR)+\mu J_{m+1}(\eta R)
J_{m}(kR)}{J_{m}(\eta R)H^{(1)}_{m+1}(kR)-\mu J_{m+1}(\eta R)
H^{(1)}_{m}(kR)}
\end {equation}
\begin {equation}\lb{bm}
b_{m}=\frac{J_{m}(kR)H^{(1)}_{m+1}(kR)-
J_{m+1}(kR)H_{m}^{(1)}(kR)}{J_{m}(\eta R)H^{(1)}_{m+1}(kR)-\mu
J_{m+1}(\eta R) H^{(1)}_{m}(kR)}.
\end {equation}

According to the Hamiltonian \eqref{eq1}, the component of the
current density is
${\vec j}=\psi^{\dag}{\vec \sigma} \psi$ where inside the quantum
dot $\psi=\psi_{t}$ and outside  $\psi=\psi_i+\psi_r $. The radial
component of the current reads as
\beq
j_{r}=\vec{j}\cdot
\vec{e}_{r}= \psi^{\dag}\left (\sigma_{x} \cos\phi+\sigma_{y}
\sin\phi\right)\psi
\eeq
or equivalently
\begin{equation}
\label{eqjrr}
 j_r =\psi^{\dag}
\begin{pmatrix}
0 & \cos\phi-i\sin\phi \\ \cos\phi+i\sin\phi & 0 \\
\end{pmatrix} \psi.
\end{equation}
Thus,
 the radial current for the reflected wave takes the
form
\begin{eqnarray}\lb{eqjrrr}
j^{r}_{r}=\frac{1}{2}\sum^{m=\infty}_{m=0}A_m(kr)
\begin{pmatrix}0 & e^{-i\phi} \\e^{-i\phi} & 0
\\\end{pmatrix}  \sum^{m=\infty}_{m=0}B_m(kr)
\end{eqnarray}
where different coefficients are given by
\begin{eqnarray}
A_m(kr) &=& (-i)^{m+1}\left[iH^{(1)\ast}_{m}(kr)
\begin{pmatrix} a^{\ast}_{m}e^{-im\phi},& a^{\ast}_{-(m+1)}e^{im\phi}\\\end{pmatrix}\right.\\
&&+ \left.
H^{(1)\ast}_{m+1}(kr)\begin{pmatrix}
a^{\ast}_{-(m+1)}e^{i(m+1)\phi},&
a^{\ast}_{m}e^{-i(m+1)\phi}\\\end{pmatrix}\right] \nonumber\\
B_m(kr) &=& i^{m+1}\left[-iH^{(1)}_{m}(kr)
\begin{pmatrix} a_{m}e^{im\phi}\\a_{-(m+1)}e^{-im\phi}\\\end{pmatrix}+
H^{(1)}_{m+1}(kr)\begin{pmatrix} a_{-(m+1)}e^{-i(m+1)\phi}\\
a_{m}e^{i(m+1)\phi}\\\end{pmatrix}\right].
\end{eqnarray}
The asymptotic behavior of the Hankel function of the first kind
for $kr\gg 1$, gives the approximate function
\begin{eqnarray}
H_m(kr)\simeq \sqrt{\frac{2}{\pi
kr}}e^{i\left(kr-\frac{m\pi}{2}-\frac{\pi}{4}\right)}
\end{eqnarray}
leads to a reduced form of \eqref{eqjrrr}
\begin{equation}\lb{jrr2}
j^{r}_{r}(\phi)=\frac{4}{k \pi
r}\sum^{m=\infty}_{m=0}|c_m|^2\left[\cos(2m+1)\phi+1\right]
\end{equation}
where we have defined
$|c_m|^2=\frac{1}{2}({|a_m|^2+|a_{-(m+1)}|^2})$. This reflected current density will
be used to determine two interesting quantities.

Let us investigate some interesting quantities related to our system and underline
their basic features. Indeed
we
can use \eqref{jrr2}
in the limit $kr\longrightarrow \infty$ to calculate the
scattering cross section $\sigma$
defined by \beq \sigma=I_r^r/(I^i/A_u)\eeq
 where $I_r^r$ is the
total reflected flux through a concentric circle and $I^i/A_u$ is
the incident flux per unit area.
From our results, we show that
$I_{r}^{r}$ takes the form
\beq
I^{r}_{r}=\int_0^{2\pi} J^{r}_{r}(\phi)r
d\phi=\frac{8}{k}\sum^{m=\infty}_{m=0}|c_m|^2
\eeq
while for the incident wave \eqref{eq.9}, we end up with $I^i/A_u=1$.
To go deeply in our study for the scattering problem for a plane
Dirac electron for different size of the circular quantum dot, we
analyze the scattering efficiency $Q$. This  is given
as the ratio between
the scattering cross section and the geometric cross section
 \beq
Q=\frac{\sigma}{2R}=\frac{4}{kR}\sum^{m=\infty}_{m=0}|c_m|^2. \eeq

Having settled the scattering efficiency and the radial current,
we proceed next  to numerically compute these quantities
in terms  of different physical parameters of our system.
This will help us to understand the effect of the energy gap and the
dot radius on the scattering in the quantum dot.

\section{Results and discussions}

To allow for a suitable interpretation of the scattering cross
section we have defined the scattering efficiency $Q$, which will
be numerically computed
under various conditions.
{Before doing so, 
we  define  different scattering
regimes. Indeed, according to the electron energy $E$ being less or above $V_{\pm}$, we
define three regimes refereed to  $E <V_-$, $V_- < E < V_+$
and $V_+<E$. Note that, 
the second regime
is a consequence of the
introduction of an energy gap inside the dot. This is in contrast
with the case of gapless graphene quantum dot \cite{29Schulz15}
where there is only two regimes. }

\begin{figure}[!ht]
\centering
\includegraphics[width=5.5cm, height=4.7cm]{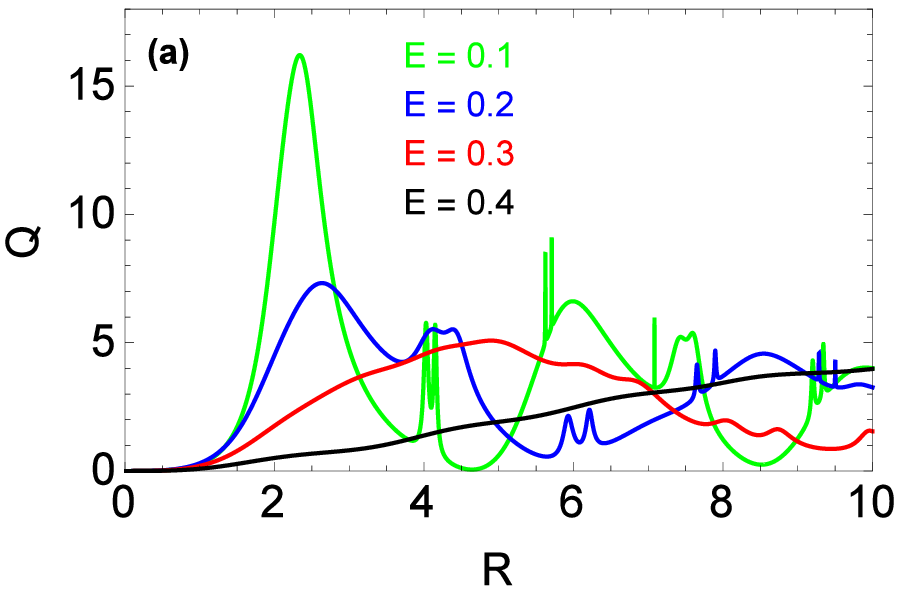}
\ \includegraphics[width=5.5cm, height=4.9cm]{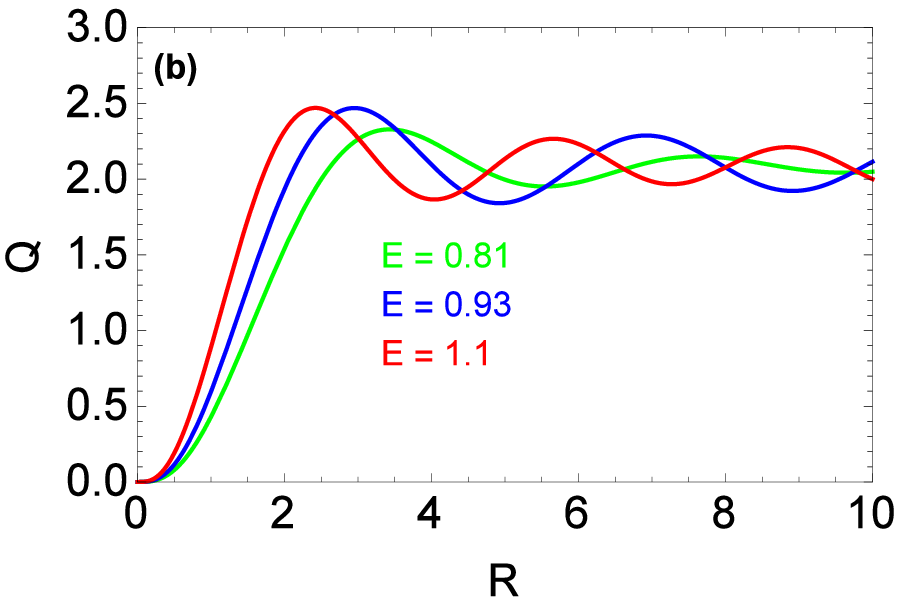}
 \  \includegraphics[width=5.5cm, height=4.9cm]{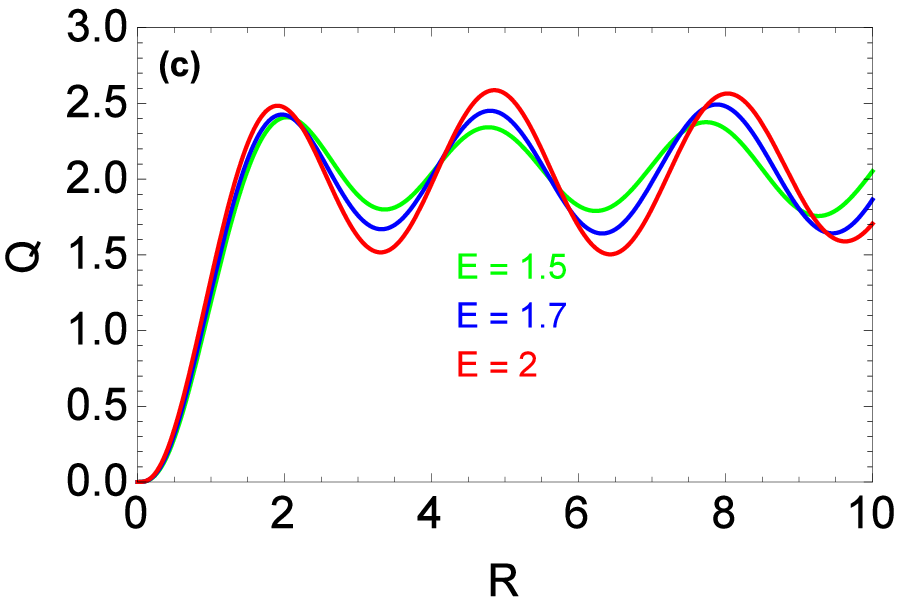}
\caption{\sf{Scattering efficiency $Q$ versus the quantum dot
radius $R$ for different values of the incident energy $E$, with
the potential $V = 1$ and gap $\Delta = 0.2$. (a): $E< V_-$, (b):
$V_-<E< V_+$, (c): $V_+<E$.}}\label{Fig2}
\end{figure}

{Numerical results for the scattering efficiency $Q$ versus the quantum dot radius $R$
for different values of the incident energy $E$, with  some choices  of the potential height $V$ and the energy gap $\Delta$,
are shown in Figure \ref{Fig2} for three different scattering regimes. 
Figure \ref{Fig2}(a) corresponds to $E<V_-$,
in this regime the region outside the dot of radius $R$ contains electrons in the conduction
band, whereas the region inside the dot contains holes in the valence band
where there is 
only evanescent
waves. From this Figure, it is clearly seen that, for small energy, when $R$ is small (close to 0), $Q$ is still null. By
increasing the dot radius, $Q$ shows an oscillatory behavior where the amplitude decreases by increasing $R$ and sharp peaks
emerge. However, by increasing $E$ the oscillations become relatively smooth. Moreover, the
scattering resonances appear, which are due to the excitation of normal modes in the quantum dot.}
{We present in Figure \ref{Fig2}(b) the
scattering efficiency $Q$ versus the quantum dot radius $R$ for
the electronic state inside the quantum dot around the Dirac point
where $V_-<E<V_+$. We observe that when the dot radius is close to
0, $Q$ is null. 
Note that in this regime there
are no available states inside the dot. In addition, by increasing $R$, the
scattering efficiency $Q$ increase almost linearly up to a
specific value of $R$ above which an oscillatory behavior sets in. The
amplitude of the oscillations decreases by increasing $E$.
However, for larger $R$ these oscillations are relatively damped.
For the third regime, the results are shown
in Figure \ref{Fig2}(c) where  the regions outside and
inside the dot have electrons in the conduction band. From
this Figure, we can see clearly that when $R$ is close to zero the
results are similar to those corresponding to $V_-<E<V_+$. But, by
increasing $R$ and for the three values of the energy, the three
curves are superimposed and increase linearly up to a specific
value of $R$ then $Q$ shows an oscillatory behavior. The amplitude
of the oscillations depends on the values of the energy $E$, it
increases as long as $E$ increased.}

\begin{figure}[!ht]
\centering
\includegraphics[width=7.4cm, height=5cm]{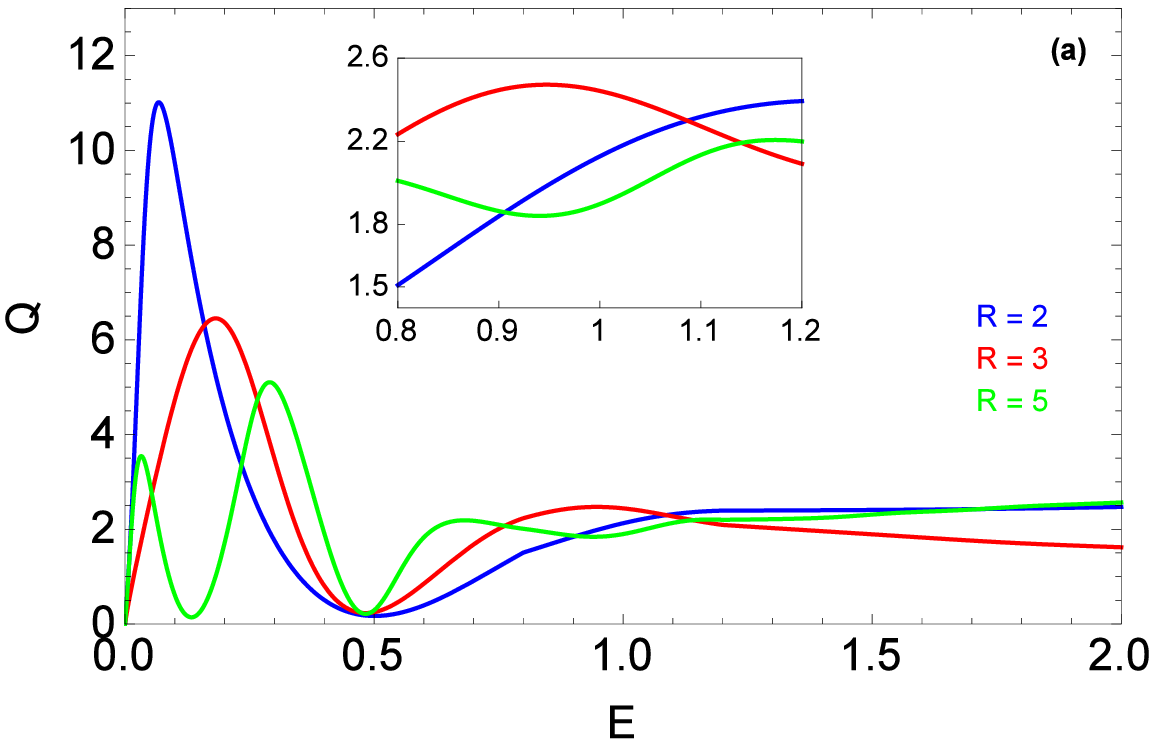} \ \
\includegraphics[width=7.4cm, height=5cm]{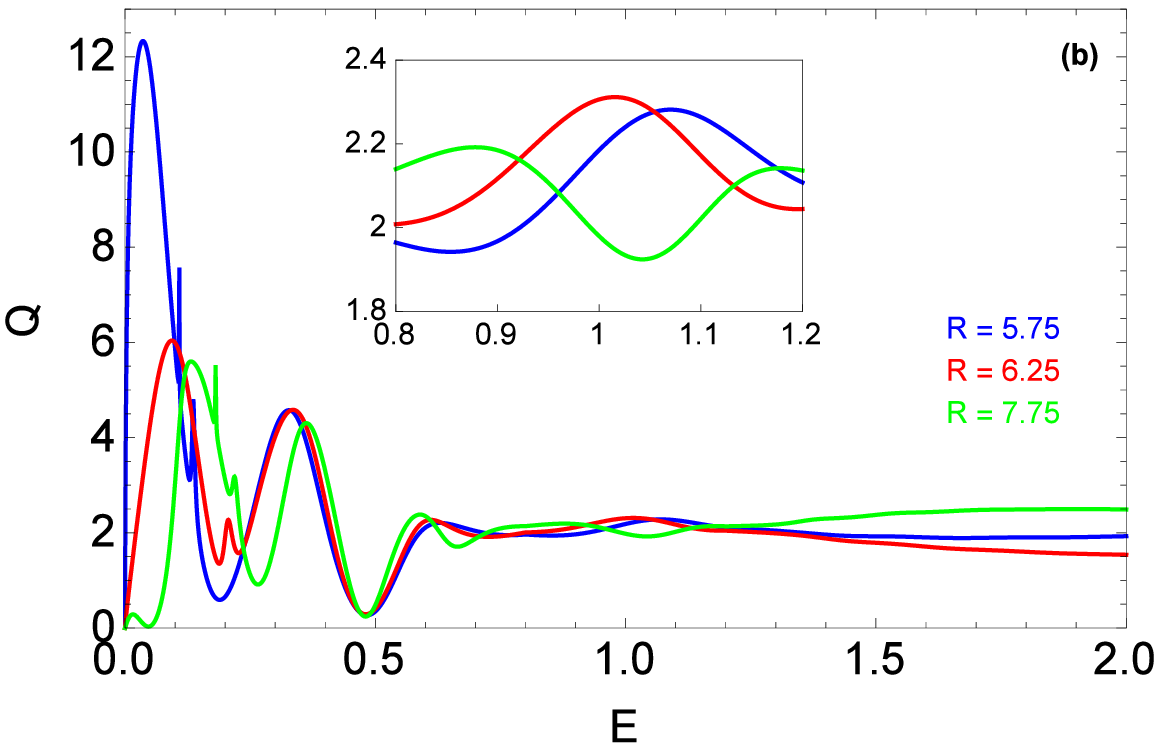}  \
\caption{\sf{Scattering efficiency $Q$ versus the energy $E$ of
the incident electron for different size of the quantum dot radius
$R$, with $\Delta=0.2$ and $V=1$. (a): $R=2, 3, 5$, (b):
$R=5.75, 6.25, 7.75$. }}\label{Fig3}
\end{figure}

{To further analyze the scattering in the three
regimes, we plot in Figure \ref{Fig3} the scattering efficiency Q
as function of incident energy $E$ for different size of the
quantum dot radius $R$. In the first regime ($E<V_-$), we can
clearly see that for small values of $R$, $Q$ is zero for $E=0$
and by increasing $E$, $Q$ shows broad maxima. The maximum of $Q$
decreases as long as $R$ is increased.
However for large $R$, $Q$ also show broad maxima and we observe
the appearance of  peaks emerging with damped oscillations. These
sharp peaks are due to the resonant excitation of normal modes of
the quantum dot,  
which are presents even if
$\Delta=0$ \cite{29Schulz15}. The results for the second regime
($V_-<E<V_+$) are shown in the inset of Figure \ref{Fig3} where 
$Q$ shows an oscillatory behavior with
small amplitude. In the third regime, we show that the oscillations
are damped and $Q$ remains constant even if $E$ increases.}\\

\begin{figure}[!ht]
\centering
\includegraphics[width=8.2cm, height=5.6cm]{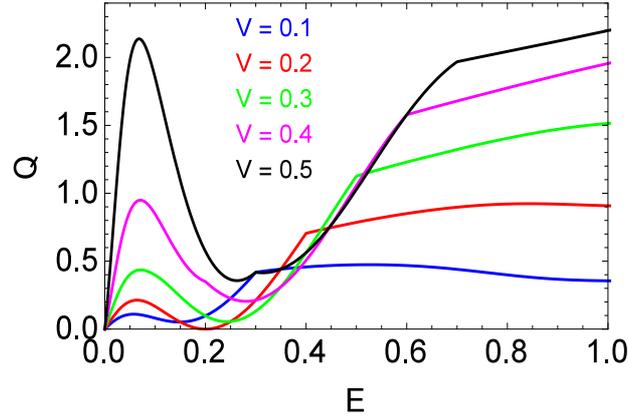}
\caption{\sf{Scattering efficiency $Q$ versus the energy E of the
incident electron for different values of the potential height
($V=0.1, 0.2, 0.3, 0.4, 0.5$) with $\Delta=0.2$ and
$R=3$.}}\label{Fig31}
\end{figure}

In order to show how the potential $V$ affects the scattering
efficiency, we plot in Figure \ref{Fig31} $Q$ as function of the
energy for different values of the potential $V$ with $\Delta=0.2$
and $R=3$.
By increasing $E$, Q shows broad maxima, which depend on the value
of 
 $V$. In fact, by increasing $V$, the maxima
increase. For large values of $E$, Q undergoes an almost linear
increase, specially when $E>V+\Delta$.

\begin{figure}[!ht]
\centering
\includegraphics[width=8.2cm, height=5.6cm]{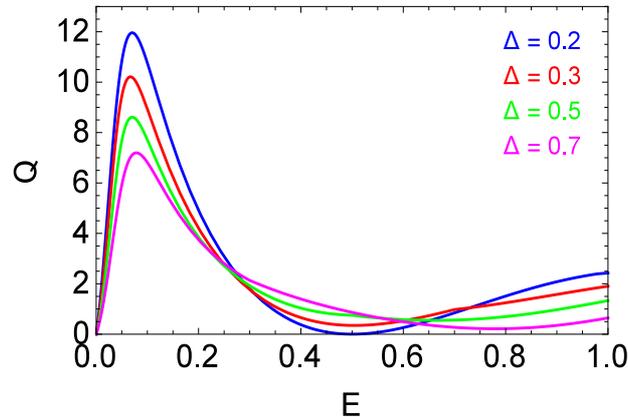}
\caption{\sf{Scattering efficiency $Q$ versus the energy E of the
incident electron for different values of the gap ($\Delta=0.1,
0.3, 0.5, 0.7$) with $V=1$ and $R=2$.}}\label{Fig32}
\end{figure}

Figure \ref{Fig32} shows the scattering efficiency $Q$ as a
function of the energy $E$. It has been performed using $V=1$, $R=2$ and taking different values for the
 gap ($\Delta=0.1, 0.3, 0.5, 0.7$). We notice that for
$E=0$, 
$Q$ is
zero whatever the values of $\Delta$ and $R$. By
increasing $E$, Q increases  until it reaches a maximum value with
different amplitudes, then decreases to a minimum value and starts
to increase again. By increasing $\Delta$, we observe that the
maxima decrease and when $E>V+\Delta$ 
$Q$ remains
constant even if the energy $E$ increases.\\

\begin{figure}[!ht]
\centering
\includegraphics[width=5.5cm, height=4.7cm]{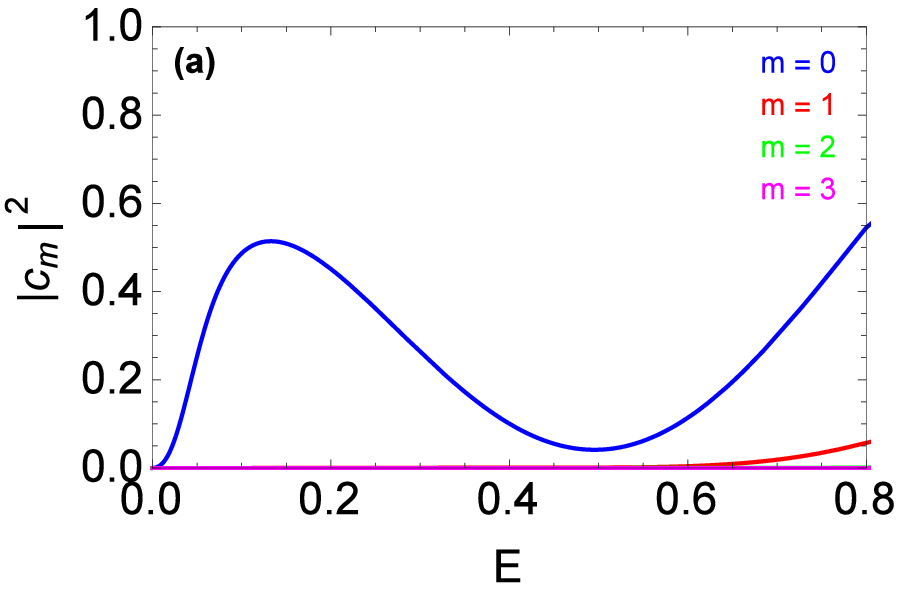}\  \includegraphics[width=5.5cm, height=4.7cm]{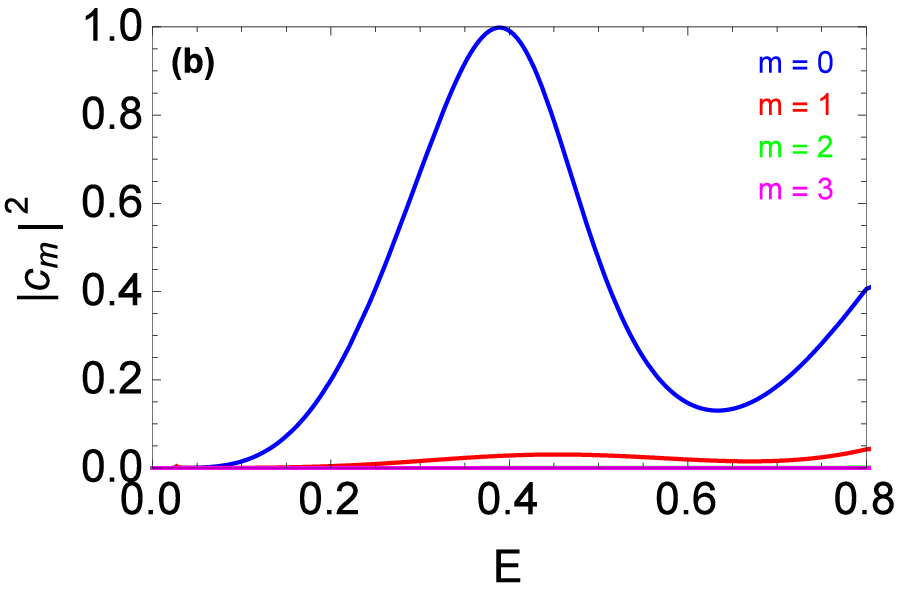}
 \ \includegraphics[width=5.5cm, height=4.7cm]{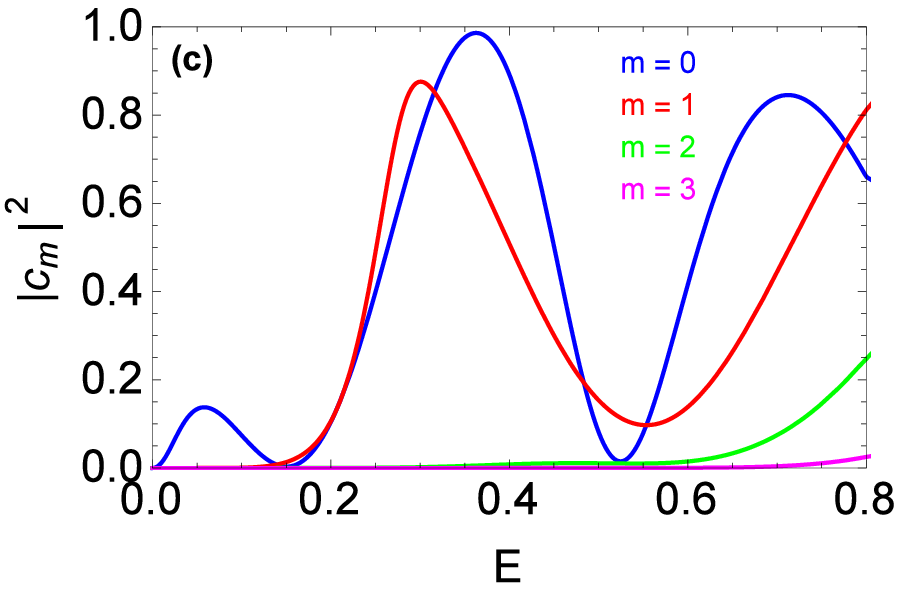}\\
\includegraphics[width=5.5cm, height=4.7cm]{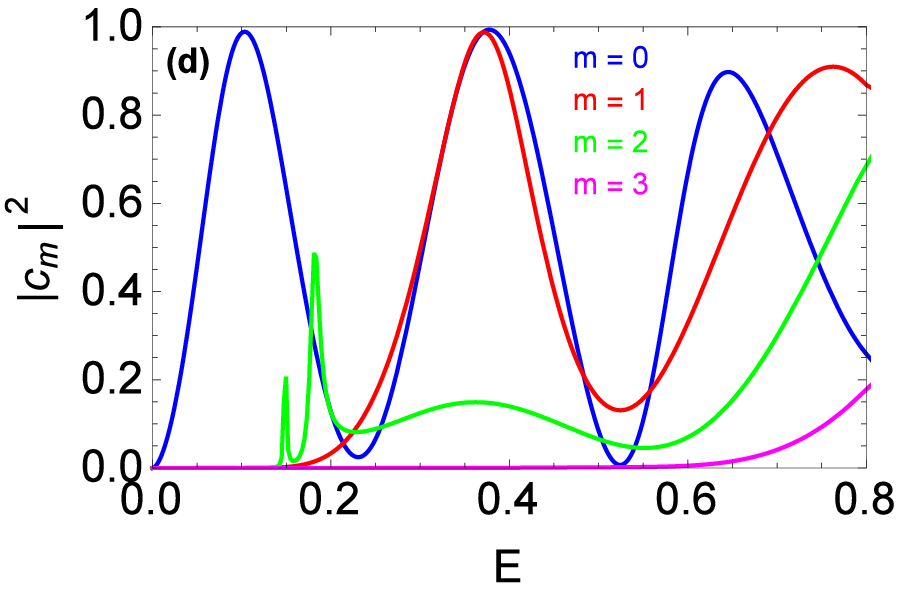}\ \includegraphics[width=5.5cm, height=4.7cm]{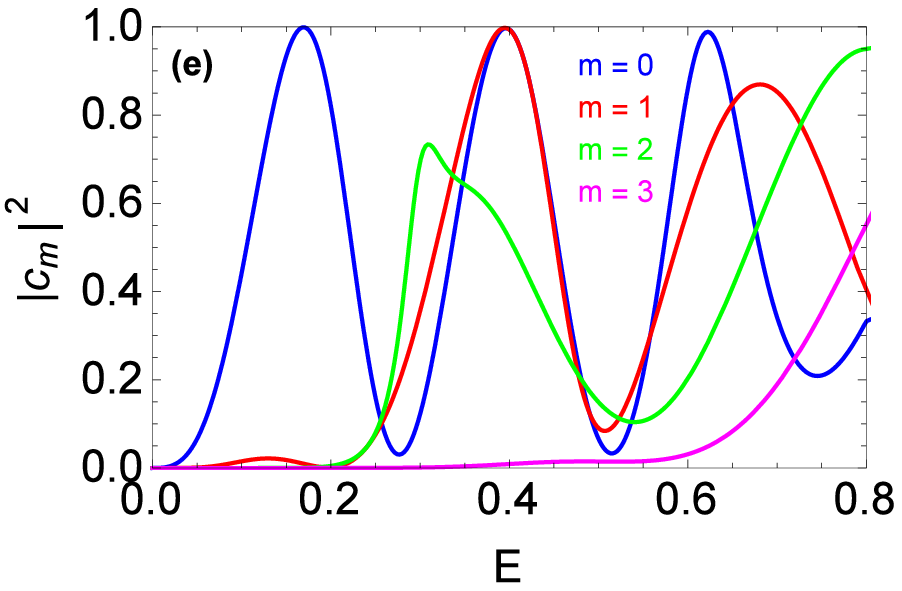}
\ \includegraphics[width=5.5cm, height=4.7cm]{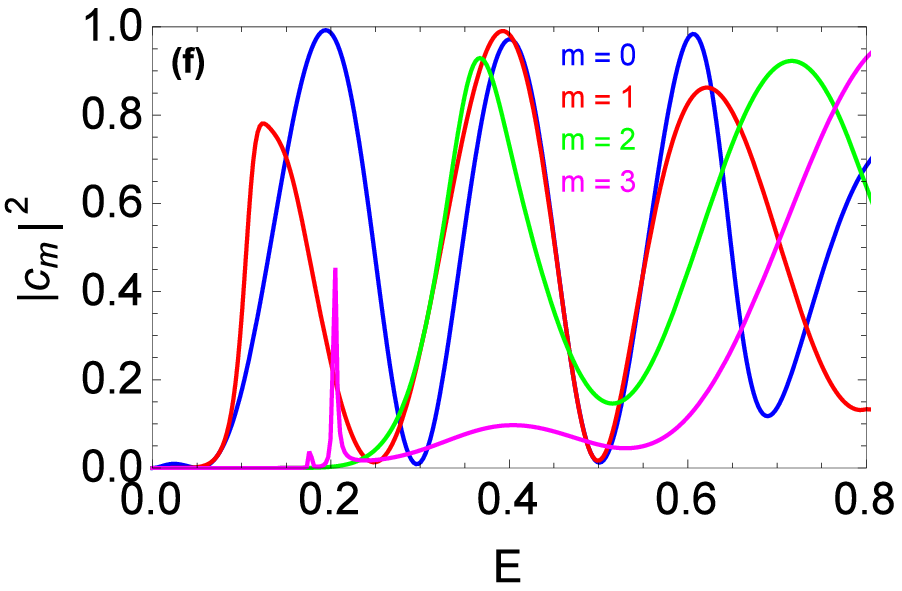}\\
\caption{\sf{Square modulus of the scattering coefficients $|
c_m|^2$ for $m=0, 1, 2, 3$ versus the energy $E$, with $V = 1$,
$\Delta=0.2$ and for six values of the quantum dot radius $R$:
$R=2$ panel (a), $R=4$ panel (b), $R=5$ panel (c), $R=6$ panel
(d), $R=7$ panel (e) and $R=7.75$ panel (f)}.}\label{Fig4}
\end{figure}

In Figure \ref{Fig4}, we plot the square modulus of the scattering
coefficients $|c_m|^2$ for $m=0, 1, 2, 3$ as function of the
energy $E$, for $V = 1$, $\Delta=0.2$ with  (a): $R=2$, (b):
$R=4$, (c): $R=5$, (d): $R=6$, (e): $R=7$, (f): $R=7.75$. From
these Figures, we observe that for zero or close to zero energy all
scattering coefficients are zero except the one corresponding to
$m=0$. By increasing $E$, we can clearly observe the contribution of
the scattering coefficients of higher orders \emph{i.e.} $m=1, 2,
3$.
By increasing $E$,  $|c_m|^2$ restores an oscillatory behavior. As compared
to the results for zero gap \cite{29Schulz15}, we notice that the
presence of an gap increases the number of oscillations.
Moreover, one can see that for some values of $E$, $|c_m|^2$
presents sharp peaks. These resonances associated with normal modes of the
quantum dot lead to the existence of sharp peaks
in Figure \ref{Fig3}, which is similar to that observed for zero
gap \cite{29Schulz15}.

\begin{figure}[!ht]
  \centering
\includegraphics[width=6.5cm, height=5.2cm]{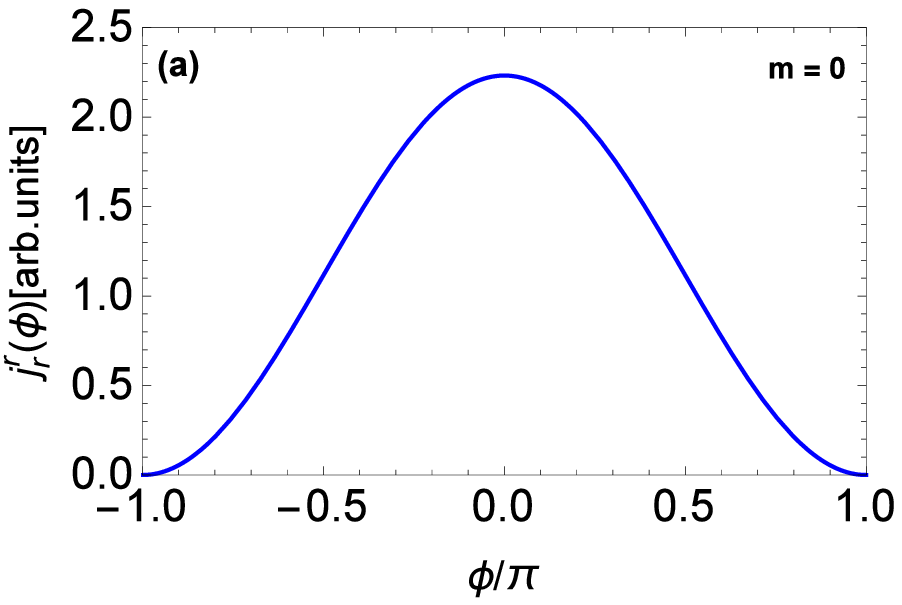}\ \ \includegraphics[width=6.5cm, height=5.2cm]{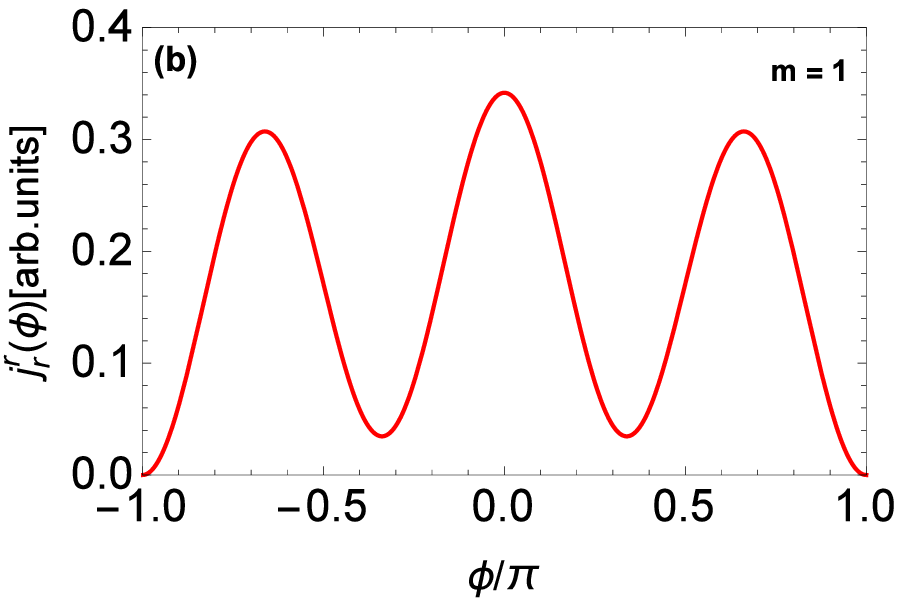}\\
\includegraphics[width=6.5cm, height=5.2cm]{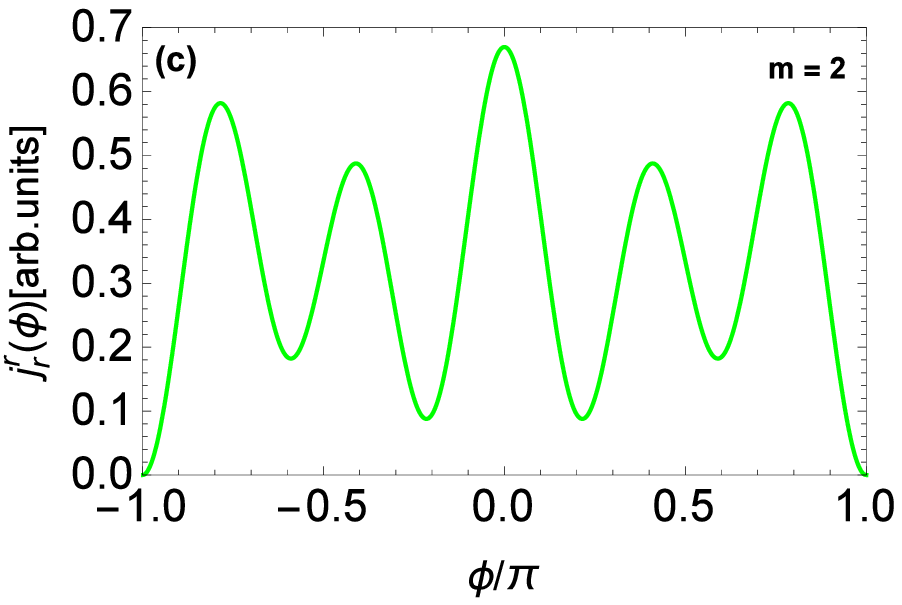}\ \ \includegraphics[width=6.5cm, height=5cm]{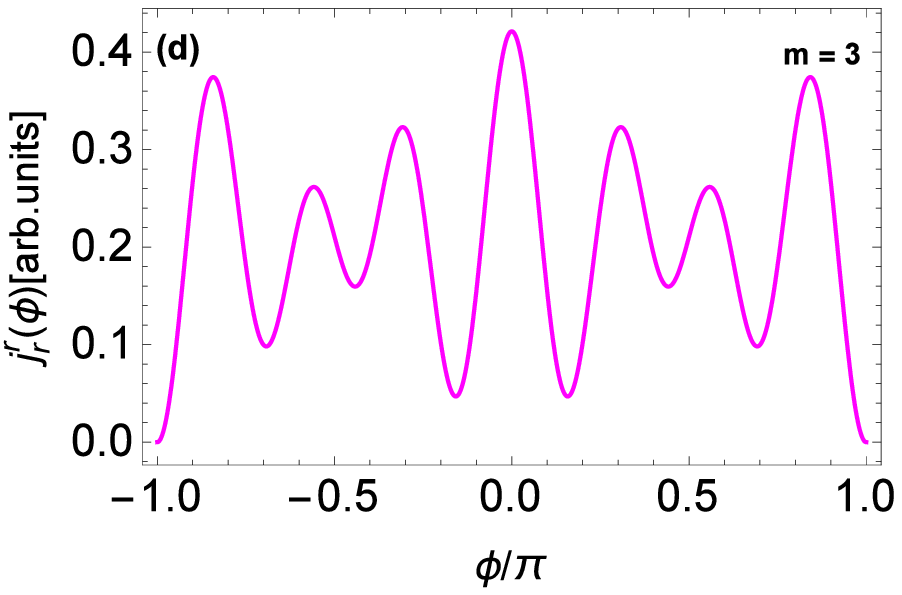}
\caption{\sf{Radial component of the far-field scattered current
$j^r_r$ as a function of the angle $\phi$ for $\Delta=0.2$ and
$V=1$. (a): $E=0.0704$ and $R=3$, (b): $E=0.484$ and $R=4$, (c):
$E=0.67$ and $R=7.75$, (d): $E=0.99$ and $R=6.25$. }}\label{Fig5}
\end{figure}

The current is defined by $j=\psi^{\dag} \sigma \psi$, where
$\psi=\psi_i + \psi_r$ outside and $\psi=\psi_t$ inside the gated
dot region. As a result of that the far-field radial component of
the reflected current $j^{r}_{r}(\phi)$ characterizes the angular
scattering is given by
\begin{equation}
j^{r}_{r}(\phi)\sim|c_m|^2\left[\cos\left((2m+1)\phi\right)+1\right].
\end{equation}

In Figure \ref{Fig5}, we plot the angular characteristic of the
reflected radial component as a function of the angle $\phi$. It
shows that only forward scattering is preferred (no
backscattering) when $\varphi=\pm \pi$. In addition, for the mode
$c_0$ (panel (a)) only forward scattering is favored. While for
higher modes more preferred scattering directions emerge. Thereby,
for $m=1$ (panel (b)) three preferred scattering directions.
However, for $m=2$ (panel (c)) five preferred scattering
directions and for $m=3$ (panel (d)) seven preferred scattering
directions. In general, each mode has $(2m + 1)$ preferred
scattering directions observable but with different amplitudes. For
small electron energies the mode ($m=0)$ is relatively broad
compared to the sharp resonances of higher modes. Resonant
scattering through one of the normal modes is also reflected in
the electron density profile in the vicinity of the quantum dot.

\section{Conclusion}

We have studied the scattering problem of an electron plane wave on a
circular electrostatically confined quantum dot in monolayer graphene with gap and compared
our results with those obtained for zero gap situation \cite{29Schulz15}.
Different scattering regimes were investigated as a function
of the radius $R$ of the quantum dot,  electrostatic potential $V$,  energy gap
$\Delta$ and  incident electron energy $E$.
We have found that scattering efficiency $Q$, for $E>V_+$ increases with increasing $R$,
first almost linearly up to a specific value of $R$ then showed an
oscillatory behavior. The amplitude of the oscillations increased
with increasing $E$. When $V_-<E<V_+$, $Q$ showed the same behavior
as for $E>V_+$, but the oscillations are relatively damped.
However, for $E < V_-$, $Q$ showed an oscillatory
behavior where their amplitudes decrease by increasing $R$.
Moreover, sharp peaks emerge, which were due to the resonant
excitations of the normal modes of the quantum dot.

The scattering efficiency $Q$ was also computed numerically as a
function of the energy by choosing different values of the
potential $V$, quantum dot radius $R$ and gap $\Delta$. We have observed
 that by increasing $E$, $Q$ shows broad maxima, which
depend on the value of V.
For larger values of $E$, $Q$ undergoes an almost linear increase
specially when $E > V+\Delta$.
However, when $E > V+\Delta$, $Q$ remains constant even if the
energy $E$ increases. It has been seen that the square modulus
of $Q$ is zero in the vicinity of $E=0$, except for $m=0$ mode.
In addition, by increasing the energy the
scattering coefficients shows an oscillatory behavior. For the
angular characteristic of the reflected radial component, we found
that each mode has $(2m + 1)$ preferred scattering directions
observable with different amplitudes.

%

\section*{Acknowledgment}

The generous support provided by the Saudi Center for Theoretical
Physics (SCTP) is highly appreciated by all authors. HB and AJ acknowledges partial support
by King Fahd University of petroleum and minerals under the theoretical physics
research group project RG171007-1 and -2.

\end{document}